\documentclass[a4paper,11pt]{article}\usepackage[]{graphicx}\usepackage[]{color}
\makeatletter
\def\maxwidth{ %
  \ifdim\Gin@nat@width>\linewidth
    \linewidth
  \else
    \Gin@nat@width
  \fi
}
\makeatother

\definecolor{fgcolor}{rgb}{0.345, 0.345, 0.345}
\newcommand{\hlkwb}[1]{\textcolor[rgb]{0.69,0.353,0.396}{#1}}%

\usepackage{framed}
\makeatletter
 {\par\unskip\endMakeFramed%
 \at@end@of@kframe}
\makeatother

\definecolor{shadecolor}{rgb}{.97, .97, .97}
\definecolor{messagecolor}{rgb}{0, 0, 0}
\definecolor{warningcolor}{rgb}{1, 0, 1}
\definecolor{errorcolor}{rgb}{1, 0, 0}
\newenvironment{knitrout}{}{} 

\usepackage{alltt}

\definecolor{myciteclr}{rgb}{0.8,0.3, 0.0}
\definecolor{mylinkclr}{rgb}{0.0,0.3,0.8}
\definecolor{mytxtclr2}{rgb}{0.4,0.25,0.0}
\usepackage{dcolumn} 
\usepackage{amsmath}
\usepackage[a4paper,vmargin=1in,includefoot]{geometry}
\usepackage[colorlinks,linkcolor=mylinkclr,citecolor=myciteclr]{hyperref}
\usepackage{rotating}
\usepackage{longtable}
\usepackage{array}
\usepackage{mathpazo}
\usepackage{booktabs}

\usepackage{natbib}
\title{Indian urban workers' labour market transitions}
\author{Jyotirmoy Bhattacharya \thanks{Ambedkar University Delhi, Email:\texttt{jyotirmoy@jyotirmoy.net}. I am grateful to seminar participants at Centre for Sustainable Employment (Azim Premji University) and Ambedkar University Delhi and to Ishan Anand and Anamitra Roychowdhury for very helpful comments. Replication code for the paper is available at: \url{https://github.com/jmoy/plfs-jb}}}
\date{November 13,2021}
\IfFileExists{upquote.sty}{\usepackage{upquote}}{}
\begin{document}
\maketitle
\begin{abstract}
This paper studies gross labour market flows and determinants of labour market transitions for urban Indian workers using a panel dataset constructed from Indian Periodic Labour Force Survey (PLFS) data for the period 2017--18 to 2019--20. Longitudinal studies based on the PLFS have been hampered by data problems that prevent a straightforward merging of the 2017--18 and 2018--19 data releases. In this paper, we propose and validate a matching procedure based on individual and household characteristics that can successfully link almost all records across these two years. We use the constructed data set to document a number of stylised facts about gross worker flows and to estimate the effects of different individual characteristics and work histories on probabilities of job gain and loss. 
\end{abstract}

\section{Introduction}

The movement of workers from job to job and from employment to non-employment and back, and the simultaneous creation and destruction of jobs by firms, are the mechanisms through which labour markets reallocate resources in response secular changes and shocks. How well they do this --- in terms of efficiently matching workers to jobs with the least amount of resources lost in waiting and search --- is an important determinant of individual welfare and aggregate productivity.

For this reason, the study of labour market transitions and gross worker and job flows has been an important part of empirical labour economics for long. Some of the key contributions in this literature are \citet{abowd1985estimating}, \citet{blanchard1990cyclical}, \citet{davis1992gross}, \citet{davis1998job}, \citet{shimer2005cyclical}, \citet{hall2005job}, \citet{fujita2009cyclicality}, \citet{shimer2012reassessing}, \citet{elsby2013unemployment}, \citet{hall2019job} and \citet{ahn2021measuring}. A recent review of evidence is provided by~\citet{davis2006flow}. This empirical literature developed in dialogue with theoretical work on search and matching models of the labour market,surveyed for example in~\citet{rogerson2011search}.

The picture of the labour market that emerges from these studies is one of a constant churning in which firms create and destroy jobs in response to growth or shrinkage in their activity and workers move from job to joblessness and other jobs either because their old job has been destroyed, or because of individual-level shocks, or in hopes of better opportunities. The changes observed in aggregate employment rates are just the small net surface result of this incessant process of reallocations at the microeconomic level. The wealth of empirical evidence on these flows that has been available for developed economies informs current research on the business cycle behaviour of the labour market and the normative and welfare impacts of different labour market institutions and policies.

Unfortunately, for long it was not possible to study these processes for the Indian economy because of the unavailability of any sources of panel data on workers. However, in just the last couple of years, such data has finally become available in the form of the official Periodic Labour Force Survey (PLFS) conducted by the Indian National Sample Survey Office (NSSO) as well as the private Consumer Pyramids Household Survey (CPHS) of the Centre for Monitoring Indian Economy(CMIE). 

The present study is part of the emerging literature that uses these sources to study labour market transitions in India. We use data from all three available annual microdata releases of the PLFS, viz. from 2017--18, 2018--19 and 2019--20, to look at gross worker flows for urban India. The paper is restricted to urban workers since the PLFS has a panel structure only for urban workers.

We begin by constructing a combined panel dataset from the three annual data releases. This is a non-trivial task because of an undocumented change in sampling unit identifiers that makes it impossible to use the provided household and individual identifiers to merge the data from the 2017--18 and 2018--19 files. This is perhaps the reason why PLFS data has not been used much for longitudinal studies so far. We overcome this difficulty by developing a matching procedure based on individual and household characteristics that can match records between these two files with a high degree of accuracy. This procedure is described in section~\ref{sec:merging}.

In section~\ref{sec:gross-flows} we use this constructed dataset to compute gross flow rates across different labour market states and record a set of stylised facts that emerge. These facts mostly match what is observed in the case of other countries, but with some specificities as well, the most important being a large difference between the patterns for men and women and the existence of large flows between salaried employment, casual employment and self-employment.

Section~\ref{sec:ind-region} documents large differences in gross flow rates between different industries, different forms of employment and different states (provinces). It finds that while most of the differences between states can be explained by differences in the composition of economic activity, there still exists for some states an unexplained state-specific effect.

Finally, section~\ref{sec:gain-loss} looks at the determinants of job gains and losses at the individual level. The results once again show a vast difference between men and women in the probabilities of job gains and losses and in the impact of marriage and child-bearing on these probabilities. The results also show differences in labour market outcomes by age and level of education and a significant impact of employment histories on transition probabilities.

\section{Related work}

As we have mentioned, due to the paucity of data, there is very limited prior research on gross flows in Indian labour markets.

\citet{sarkar2019employment} used Indian Human Development Survey (IHDS) panel data from 2005 and 2012 to look at labour market transitions of Indian women in the context of the debate around the low and decreasing Labour Force Participation Rate (LFPR) of Indian women. They found that women have higher exit probabilities from and lower entry probability into employment compared to men. They also studied different determinants of entry and exit. The key limitation of their paper, imposed by their data source, was the long gap of seven years between the two observations. As a result they faced serious attrition problems with 20\% of women in the 2005 sample not being present in the 2012 sample. Also, they could not observe labour market movements at higher frequencies, which we find to be significant.

\citet{deshpande21} use  Consumer Pyramids Households Survey data from 2016--2019 to look at women's labour market transitions, once again in the context of the women's LFPR debate. The panel design of the CPHS allows them to track individuals over the entire study period and one of their key findings, corroborated by the present study, is that the low cross-sectional labour participation rate of women does not truly capture the work histories of women. They find large gross movements of women into and out of work at a quarterly frequency. Though the the average cross-sectional labour force-participation rate of women was only 14.6\% in their data, the percentage of women who were observed to be employed at least once in the entire study period was much higher at 44\%.

The paper closest to the present one is that of~\citet{menon21} who look at urban labour market transitions and gross flows using the same Periodic Labour Force Survey data that we use . However they only work with the 2017--18 and 2018--19 data releases and do not address the coding change between these two data releases. They are therefore forced to split worker histories at the boundary of these two years. From each year, they consider only one panel --- that which started in the first quarter in that year --- and only look at transitions at an annual frequency. They also look at only three broad labour market states --- employment, unemployment and non-participation. In contrast our recoding algorithm allows us to merge data from 2017--18 and 2018--19 files, which we further combine with the 2019-20 files to create one combined data set that includes all the PLFS data released so far. As a result we work with much larger sample sizes. We look at all the panels and calculate transition rates at a quarterly frequency with a finer set of labour market states which distinguishes between salaried work, casual work and self-employment. As a result we are able to pick out gross flows at a higher granularity that brings out more clearly the dynamic nature of the urban Indian labour market.

\section{Data}

\subsection{Survey design}
The Periodic Labour Force Survey (PLFS) is a quarterly survey on employment issues conducted by the National Sample Survey Office (NSSO) of the Government of India. Started in 2017-18, it replaces the earlier quinquennial employment-unemployment surveys of the NSSO. Annual reports and microdata for the survey are published in a July-June cycle. So far data has been released for the years 2017--18, 2018--19 and 2019--20. The design of the survey is documented in \citet{plfs16instr}.

The PLFS covers both rural and urban areas. In urban areas it has a rotational panel design, with a new panel starting every quarter and being visited for four successive quarters. This is the first official household survey in India with a panel structure and the primary aim of this paper is to use this structure to study transition processes in the labour market that cannot be captured by purely cross-sectional data. Since rural households are visited only once, this paper is restricted to urban households.

The PLFS is a stratified, multi-stage survey. We describe the sample design only for the urban areas. The first-stage units (FSUs) are urban blocks from the Urban Frame Survey of the NSSO. This sampling frame changes every two years, so in the period of our study one sampling frame is used for 2017--18 and 2018--19 and another for 2019--20.

\begin{knitrout}
\definecolor{shadecolor}{rgb}{0.969, 0.969, 0.969}\color{fgcolor}\begin{table}

\caption{\label{tab:panel-samp-size}Number of households visited in each panel in each quarter}
\centering
\resizebox{\linewidth}{!}{
\fontsize{10}{12}\selectfont
\begin{tabular}[t]{lrrrrrrrrrr}
\toprule
\multicolumn{11}{c}{No. of households visited (thousands)} \\
\cmidrule(l{3pt}r{3pt}){1-11}
\multicolumn{1}{c}{} & \multicolumn{10}{c}{Panel identifier} \\
\cmidrule(l{3pt}r{3pt}){2-11}
Year-Quarter & P11 & P12 & P13 & P14 & P15 & P16 & P17 & P18 & P21 & P22\\
\midrule
2017-Q3 & 11.4 &  &  &  &  &  &  &  &  & \\
2017-Q4 & 11.1 & 11.5 &  &  &  &  &  &  &  & \\
2018-Q1 & 11.0 & 11.2 & 11.5 &  &  &  &  &  &  & \\
2018-Q2 & 10.8 & 11.0 & 11.2 & 11.5 &  &  &  &  &  & \\
\addlinespace
2018-Q3 &  & 11.0 & 11.1 & 11.3 & 11.5 &  &  &  &  & \\
2018-Q4 &  &  & 11.1 & 11.1 & 11.3 & 11.5 &  &  &  & \\
2019-Q1 &  &  &  & 11.0 & 11.2 & 11.3 & 11.4 &  &  & \\
2019-Q2 &  &  &  &  & 11.2 & 11.3 & 11.4 & 11.1 &  & \\
\addlinespace
2019-Q3 &  &  &  &  &  & 10.9 & 10.9 & 10.8 & 11.4 & \\
2019-Q4 &  &  &  &  &  &  & 11.3 & 11.1 & 11.3 & 11.5\\
\bottomrule
\end{tabular}}
\end{table}

\end{knitrout}

Table~\ref{tab:panel-samp-size} gives the number of households in each panel visited in each quarter. The staggered pattern of the table demonstrates the rotating nature of the panels. The table starts with 2017-Q3 since the PLFS follows a July-June cycle. Following the official PLFS documentation, the panels are identified as $P_{ij}$ where $i$ refers to the sampling frame used and $j$ to the serial number of the panel. 

The broad outline of the survey design is as follows. A first stage of stratification is carried out by dividing towns and cities on the basis of population. From these strata, blocks, which are the first-stage units, are selected using probability proportional to size with replacement. Larger blocks are divided into sub-blocks and only two sub-blocks are selected from thse blocks for further sampling.

Households in blocks are stratified on the basis of the general education level of their members. From these second-stage strata households are selected for panels on the basis of simple random sampling without replacement.

\subsection{Initial validation}\label{sec:validation}

We merge the data files for all the three years and run the following validation checks to ensure that distinct households or persons are not merged together due to data errors in the original data.

\begin{enumerate}
\item The religion and social group of the household should not have changed.
\item The size of the household should not have changed by more than 3.
\item A person's gender and relation to the head of the household should not have changed.
\item A person's age should not have changed by more than 4 years.\footnote{Given that each person is follwed over four quarters, age in years should not change by more than 1. However, examination of the reported age distribution shows significant bunching around multiples of 2 and 5 which suggests imprecise reporting of age. Since the purpose of this step is only to prevent incorrect linking, we keep a wider acceptance range for reported age.}
\end{enumerate}

While some of these variables can in fact change for households and persons, the frequency of such changes is rare enough that most reported changes are likely to be due to data errors. We therefore remove all the records of households which fail any of these tests, removing the records of the entire household in case any of its members fails any of the person-level tests. 634 households are removed from the sample as a result.

\subsection{Constructing the panel dataset}\label{sec:merging}

In the microdata files, each household is supposed to be uniquely identified by the combination of panel number, first-stage unit (FSU) number, sub-block number, second stage stratum number and sample household number. Each person within a household has a person number which remains the same across visits. In principle this should lead to a simple linking of households and individuals across different years' microdata files. However, a number of problems emerge in actually carrying out this process using the files provided by the NSSO. These data issues have also been documented by \cite{menon21} and \citet{abdul2021using}.

First, the panel identifier is not provided in the released data. We work around this by inferring panel numbers on the basis of the documented panel schedule, visit number and the year and quarter of the visit.

Second, in the 2017--18 revisit file for persons, the quarter numbers are recorded as 3, 4 or 5 instead of the expected 2, 3 and 4. Tabulating the visit numbers against the quarter numbers shows no third visits in the reported quarter 3 and no fourth visits in the reported quarter 4. Thus the most plausible explanation is that quarter 2 has been incorrectly coded as quarter 3 and so on. We correct the records with this assumption.

The third, and gravest, difficulty is that an entirely different sets of first-stage unit (FSU) numbers are used in the 2017--18 and the 2018--19 data releases even though both years are supposed to use the same sampling frame. This change is not explained in the PLFS documentation. Because of this change it is not possible to directly link data across these two years.

We develop a matching procedure to overcome this problem. We consider the development of this procedure to be a major contribution of this paper since in the absence of such a procedure it would be impossible to link together panels that began in 2017-18 and were continued into 2018-19. For these panels we would lose the opportunity to track households over the entire four-quarter span of observation.

Our procedure is based on the hypothesis, eventually confirmed, that the change in the FSU number is a simple renumbering and not a reorganization of the FSU and that within the renumbered FSUs the sub-block serial number, second-stage stratum number and sample household numbers still point to the same household.

For each state/union territory (province) and district we consider all possible (FSU number in 2017--18, FSU number in 2018--19) pairs as renumbering candidates. For each such pair we compare households identified by the same sub-block, second-stage stratum and household number and count households of size greater than two that have matching panel number, religion and social group and whose corresponding members have the same sex, relation to the head of the household and general education level. The number of households which successfully match on these criteria becomes the score of the candidate renumbering pair. We exclude households of size one or two to avoid spurious matches. The variables on which we match, specially the general education level, could presumably change over time, but the proportion of genuinely matching households with such changes is likely to be very small at a quarterly frequency. To minimize the possibility of genuine changes causing failed matches, we consider for each household only the last visit in 2017--18 and the first visit in 2018--19.

For each FSU number in 2017-18 we pick the FSU number in 2018-19 with the highest score obtained in the previous step as the best match. In case a 2018--19 FSU number is selected as the best match for more than one 2017--18 FSU number we drop all the pairs involved.

At the end of this process we are able to find best matches for 4272 out of the 4320 FSUs from 2017--18 that were to be revisited in 2018--19. 

To check on the quality of this mapping we use it to rewrite all the FSU numbers in the 2017--18 data and rerun the validation tests from section~\ref{sec:validation}. Only 62 out of 34,452 households that were to be matched fail these validation tests. This confirms our hypothesis that FSU were renumbered and not reorganized and demonstrates the procedure's success in inferring that renumbering.

Note that it is appropriate to validate the procedure by comparing variables like religion or relation to head which were used in the scoring function for FSU renumbering candidate pairs. This is because the validation is run on the entire set of households and not just on the set of matching households which were used for the scoring. Had we been mistaken in our assumption that the FSUs had only been renumbered and not reorganized or had we not correctly inferred the renumbering, the set of matching households would be very small during the scoring stage and subsequently linking all the other households using the FSU renumbering inferred would have led to a large number of households failing the validation tests.

As a further check on this procedure we run it on the 2018-19 and 2019-20 data sets in which no change in FSU numbers actually occurred. Our procedure correctly matches each FSU number in 2018-19 to itself in 2019-20, further confirming its validity.

The rest of this paper uses the data set with the 2017--18 FSU numbers replaced by the inferred 2018--19 renumbering. Households belonging the the few FSUs which could not be matched as taken as attrited at the end of 2017-18.

\subsection{Age group and study period}

All our analyses are restricted to individuals in the working age group of 15 to 65 years.

We only use data from 2017-Q3 to 2019-Q4 in order to exclude the impact of the COVID pandemic.

\subsection{Employment status}

\begin{knitrout}
\definecolor{shadecolor}{rgb}{0.969, 0.969, 0.969}\color{fgcolor}\begin{table}

\caption{\label{tab:emp-states}Employment states}
\centering
\fontsize{10}{12}\selectfont
\begin{tabular}[t]{l>{\raggedright\arraybackslash}p{20em}>{\raggedright\arraybackslash}p{7em}}
\toprule
State & Description & PLFS codes\\
\midrule
slf-emp & Self-employed in household enterprise or helper in household enterprise & 11,12,21\\
\addlinespace
csl-emp & Casual wage labour & 41, 42, 51\\
\addlinespace
sal-emp & Regular salaried/wage employee & 31\\
\addlinespace
sck-emp & Had work but did not work due to sickness & 61, 71\\
\addlinespace
nwrk & Had work but did not work due to other reasons & 62, 72\\
\addlinespace
unemp & Unemployed (not engaged in work but available for work) & 81, 82\\
\addlinespace
nopart & Not in labour force (not available for work) & 91, 92, 93, 94, 95, 97, 98, 99\\
\addlinespace
attrit & Attrited from the sample & \\
\bottomrule
\end{tabular}
\end{table}

\end{knitrout}

The PLFS collects information regarding employment status for three different reference periods---the usual principal/subsidiary status with a reference period of 365 days  prior to the visit (collected only on the first visit), daily statuses for each day of the week prior to the visit and a current weekly status derived from these daily statuses. 

All our analyses are based on the current weekly status. The usual status is not collected on revisits and is therefore not useful in a longitudinal study. 

The labour market states I use in the tabulations are obtained by combining PLFS's two-digit status codes \citep{plfs16instr} into a smaller number of categories. The list of states and their description is given in Table~\ref{tab:emp-states}.

\subsection{Attrition}

\begin{knitrout}
\definecolor{shadecolor}{rgb}{0.969, 0.969, 0.969}\color{fgcolor}\begin{table}

\caption{\label{tab:attr-by-panel}Percentage of households attrited in each visit after first}
\centering
\fontsize{10}{12}\selectfont
\begin{tabular}[t]{rrrrrrrrrr}
\toprule
\multicolumn{1}{c}{} & \multicolumn{9}{c}{\% of households attrited} \\
\cmidrule(l{3pt}r{3pt}){2-10}
\multicolumn{1}{c}{} & \multicolumn{9}{c}{Panel} \\
\cmidrule(l{3pt}r{3pt}){2-10}
Visit & P11 & P12 & P13 & P14 & P15 & P16 & P17 & P18 & P21\\
\midrule
2 & 2.6 & 2.1 & 2.5 & 1.8 & 1.5 & 1.7 & 0.0 & 2.5 & 0.7\\
3 & 4.0 & 3.6 & 3.1 & 2.9 & 2.7 & 1.6 & 3.9 & 0.2 & \\
4 & 5.3 & 4.0 & 3.8 & 3.8 & 2.6 & 4.7 & 0.4 &  & \\
\bottomrule
\end{tabular}
\end{table}

\end{knitrout}

Table~\ref{tab:attr-by-panel} gives the attrition rates. The attrition rates are comparable to other surveys of a similar nature. In fact some of the attrition rates of less than 1\% are quite unusual and surprising. 

Still, even these low attrition rates are a cause for concern since their magnitude will turn out to be comparable to the magnitude of the gross flows between labour market states which are the focus of this paper. Attrition would not make a difference if the histories of those attrited were to be similar to those not attrited. But this is unlikely to be true, since at least one source of attrition is individuals moving from their previous address and the labour market trajectories of these movers are likely to be different from those who continue to stay at their older address. 

There exists a literature on adjusting for attrition in longitudinal surveys, and in the context of gross labour flows \citet{abowd1985estimating} and \citet{feng2013misclassification} for example use such adjustments. But these adjustments require strong assumptions about the unobserved attrition process which we are not comfortable making. We therefore report attrition rates side-by-side with the estimated flow rates so that readers may judge for themselves the worst case bounds on the effects of attrition on the estimates on the lines of \citet{manski2009identification}.

\section{Gross Flows}\label{sec:gross-flows}

\subsection{Method}

In this section, we look at pairs of observations of the same individual over successive quarters to compute gross flow rates between the labour market states defined in Table~\ref{tab:emp-states}. 

For each gender and each period we form matrices of gross flows where the entry in the $i$-th row and the $j$-th column in the quarter $t$ matrix is the estimated number of individuals in labour market state $i$ in quarter $t$ who moved to labour market state $j$ in quarter $t+1$ as a percentage of the total population of that gender in that quarter. Sampling weights provided by the NSSO are used while calculating these percentages. We then average these matrices over all quarters $t$ in our study period to produce Table~\ref{tab:transmat}. The numbers in parentheses in the table are transition probabilities expressed as percentages, where the parenthesised entry in the $i$-th row and $j$-th column is number of persons in state $i$ who go to state $j$ in the next quarter, as a percentage of the total number of persons in state $i$ in the initial quarter. These probabilies are also averaged over quarters.

\begin{knitrout}
\definecolor{shadecolor}{rgb}{0.969, 0.969, 0.969}\color{fgcolor}\begin{table}

\caption{\label{tab:transmat}Occupancy of and gross flow between employment states, numbers in parentheses are transition probabilites as percentages}
\centering
\resizebox{\linewidth}{!}{
\fontsize{10}{12}\selectfont
\begin{tabular}[t]{lrrrrrrrrr}
\toprule
\multicolumn{1}{c}{} & \multicolumn{9}{c}{percentage} \\
\cmidrule(l{3pt}r{3pt}){2-10}
\multicolumn{1}{c}{} & \multicolumn{9}{c}{Emp. status [t+1]} \\
\cmidrule(l{3pt}r{3pt}){2-10}
Emp.\ status [t] & slf-emp & csl-emp & sal-emp & unemp & nopart & sck-emp & nwrk & attrit & ALL\\
\midrule
\addlinespace[0.3em]
\multicolumn{10}{l}{\textit{Female}}\\
\hspace{1em}slf-emp & 4.66 & 0.04 & 0.09 & 0.05 & 0.79 & 0.02 & 0.09 & 0.09 & 5.82\\
\textcolor{mytxtclr2}{\hspace{1em}} & \textcolor{mytxtclr2}{(79.97)} & \textcolor{mytxtclr2}{(0.66)} & \textcolor{mytxtclr2}{(1.52)} & \textcolor{mytxtclr2}{(0.92)} & \textcolor{mytxtclr2}{(13.53)} & \textcolor{mytxtclr2}{(0.29)} & \textcolor{mytxtclr2}{(1.58)} & \textcolor{mytxtclr2}{(1.52)} & \textcolor{mytxtclr2}{(100.00)}\\
\hspace{1em}csl-emp & 0.06 & 1.20 & 0.09 & 0.08 & 0.31 & 0.00 & 0.00 & 0.03 & 1.78\\
\textcolor{mytxtclr2}{\hspace{1em}} & \textcolor{mytxtclr2}{(3.44)} & \textcolor{mytxtclr2}{(67.27)} & \textcolor{mytxtclr2}{(5.11)} & \textcolor{mytxtclr2}{(4.25)} & \textcolor{mytxtclr2}{(17.66)} & \textcolor{mytxtclr2}{(0.07)} & \textcolor{mytxtclr2}{(0.28)} & \textcolor{mytxtclr2}{(1.91)} & \textcolor{mytxtclr2}{(100.00)}\\
\hspace{1em}sal-emp & 0.08 & 0.05 & 8.75 & 0.10 & 0.58 & 0.02 & 0.17 & 0.31 & 10.06\\
\textcolor{mytxtclr2}{\hspace{1em}} & \textcolor{mytxtclr2}{(0.77)} & \textcolor{mytxtclr2}{(0.52)} & \textcolor{mytxtclr2}{(86.98)} & \textcolor{mytxtclr2}{(1.02)} & \textcolor{mytxtclr2}{(5.72)} & \textcolor{mytxtclr2}{(0.17)} & \textcolor{mytxtclr2}{(1.74)} & \textcolor{mytxtclr2}{(3.09)} & \textcolor{mytxtclr2}{(100.00)}\\
\hspace{1em}unemp & 0.04 & 0.08 & 0.10 & 1.57 & 0.56 & 0.00 & 0.00 & 0.10 & 2.46\\
\textcolor{mytxtclr2}{\hspace{1em}} & \textcolor{mytxtclr2}{(1.78)} & \textcolor{mytxtclr2}{(3.31)} & \textcolor{mytxtclr2}{(4.10)} & \textcolor{mytxtclr2}{(63.60)} & \textcolor{mytxtclr2}{(22.92)} & \textcolor{mytxtclr2}{(0.01)} & \textcolor{mytxtclr2}{(0.12)} & \textcolor{mytxtclr2}{(4.16)} & \textcolor{mytxtclr2}{(100.00)}\\
\hspace{1em}nopart & 0.76 & 0.26 & 0.56 & 0.50 & 75.56 & 0.01 & 0.03 & 1.76 & 79.43\\
\textcolor{mytxtclr2}{\hspace{1em}} & \textcolor{mytxtclr2}{(0.96)} & \textcolor{mytxtclr2}{(0.32)} & \textcolor{mytxtclr2}{(0.71)} & \textcolor{mytxtclr2}{(0.63)} & \textcolor{mytxtclr2}{(95.12)} & \textcolor{mytxtclr2}{(0.01)} & \textcolor{mytxtclr2}{(0.04)} & \textcolor{mytxtclr2}{(2.22)} & \textcolor{mytxtclr2}{(100.00)}\\
\hspace{1em}sck-emp & 0.01 & 0.00 & 0.02 & 0.00 & 0.01 & 0.01 & 0.00 & 0.00 & 0.06\\
\textcolor{mytxtclr2}{\hspace{1em}} & \textcolor{mytxtclr2}{(23.93)} & \textcolor{mytxtclr2}{(0.00)} & \textcolor{mytxtclr2}{(38.05)} & \textcolor{mytxtclr2}{(1.82)} & \textcolor{mytxtclr2}{(17.68)} & \textcolor{mytxtclr2}{(11.50)} & \textcolor{mytxtclr2}{(4.99)} & \textcolor{mytxtclr2}{(2.03)} & \textcolor{mytxtclr2}{(100.00)}\\
\hspace{1em}nwrk & 0.10 & 0.00 & 0.16 & 0.00 & 0.05 & 0.00 & 0.05 & 0.01 & 0.39\\
\textcolor{mytxtclr2}{\hspace{1em}} & \textcolor{mytxtclr2}{(25.87)} & \textcolor{mytxtclr2}{(1.19)} & \textcolor{mytxtclr2}{(42.27)} & \textcolor{mytxtclr2}{(1.29)} & \textcolor{mytxtclr2}{(11.76)} & \textcolor{mytxtclr2}{(0.76)} & \textcolor{mytxtclr2}{(13.71)} & \textcolor{mytxtclr2}{(3.15)} & \textcolor{mytxtclr2}{(100.00)}\\
\hspace{1em}ALL & 5.71 & 1.63 & 9.78 & 2.31 & 77.85 & 0.05 & 0.36 & 2.31 & 100.00\\
\textcolor{mytxtclr2}{\hspace{1em}} & \textcolor{mytxtclr2}{(5.71)} & \textcolor{mytxtclr2}{(1.63)} & \textcolor{mytxtclr2}{(9.78)} & \textcolor{mytxtclr2}{(2.31)} & \textcolor{mytxtclr2}{(77.85)} & \textcolor{mytxtclr2}{(0.05)} & \textcolor{mytxtclr2}{(0.36)} & \textcolor{mytxtclr2}{(2.31)} & \textcolor{mytxtclr2}{(100.00)}\\
\addlinespace[0.3em]
\multicolumn{10}{l}{\textit{Male}}\\
\hspace{1em}slf-emp & 24.13 & 0.32 & 0.58 & 0.36 & 0.35 & 0.07 & 0.32 & 0.60 & 26.73\\
\textcolor{mytxtclr2}{\hspace{1em}} & \textcolor{mytxtclr2}{(90.27)} & \textcolor{mytxtclr2}{(1.22)} & \textcolor{mytxtclr2}{(2.16)} & \textcolor{mytxtclr2}{(1.34)} & \textcolor{mytxtclr2}{(1.31)} & \textcolor{mytxtclr2}{(0.25)} & \textcolor{mytxtclr2}{(1.21)} & \textcolor{mytxtclr2}{(2.23)} & \textcolor{mytxtclr2}{(100.00)}\\
\hspace{1em}csl-emp & 0.35 & 7.90 & 0.35 & 0.48 & 0.23 & 0.01 & 0.04 & 0.28 & 9.62\\
\textcolor{mytxtclr2}{\hspace{1em}} & \textcolor{mytxtclr2}{(3.63)} & \textcolor{mytxtclr2}{(82.06)} & \textcolor{mytxtclr2}{(3.60)} & \textcolor{mytxtclr2}{(4.98)} & \textcolor{mytxtclr2}{(2.39)} & \textcolor{mytxtclr2}{(0.06)} & \textcolor{mytxtclr2}{(0.38)} & \textcolor{mytxtclr2}{(2.90)} & \textcolor{mytxtclr2}{(100.00)}\\
\hspace{1em}sal-emp & 0.56 & 0.29 & 30.07 & 0.51 & 0.35 & 0.05 & 0.28 & 1.05 & 33.16\\
\textcolor{mytxtclr2}{\hspace{1em}} & \textcolor{mytxtclr2}{(1.69)} & \textcolor{mytxtclr2}{(0.87)} & \textcolor{mytxtclr2}{(90.69)} & \textcolor{mytxtclr2}{(1.52)} & \textcolor{mytxtclr2}{(1.04)} & \textcolor{mytxtclr2}{(0.14)} & \textcolor{mytxtclr2}{(0.86)} & \textcolor{mytxtclr2}{(3.18)} & \textcolor{mytxtclr2}{(100.00)}\\
\hspace{1em}unemp & 0.37 & 0.54 & 0.49 & 4.44 & 0.59 & 0.00 & 0.03 & 0.34 & 6.80\\
\textcolor{mytxtclr2}{\hspace{1em}} & \textcolor{mytxtclr2}{(5.50)} & \textcolor{mytxtclr2}{(7.93)} & \textcolor{mytxtclr2}{(7.22)} & \textcolor{mytxtclr2}{(65.29)} & \textcolor{mytxtclr2}{(8.70)} & \textcolor{mytxtclr2}{(0.04)} & \textcolor{mytxtclr2}{(0.37)} & \textcolor{mytxtclr2}{(4.96)} & \textcolor{mytxtclr2}{(100.00)}\\
\hspace{1em}nopart & 0.32 & 0.21 & 0.30 & 0.55 & 20.49 & 0.00 & 0.01 & 0.63 & 22.52\\
\textcolor{mytxtclr2}{\hspace{1em}} & \textcolor{mytxtclr2}{(1.44)} & \textcolor{mytxtclr2}{(0.93)} & \textcolor{mytxtclr2}{(1.33)} & \textcolor{mytxtclr2}{(2.45)} & \textcolor{mytxtclr2}{(90.97)} & \textcolor{mytxtclr2}{(0.02)} & \textcolor{mytxtclr2}{(0.06)} & \textcolor{mytxtclr2}{(2.82)} & \textcolor{mytxtclr2}{(100.00)}\\
\hspace{1em}sck-emp & 0.09 & 0.01 & 0.05 & 0.01 & 0.01 & 0.02 & 0.01 & 0.01 & 0.20\\
\textcolor{mytxtclr2}{\hspace{1em}} & \textcolor{mytxtclr2}{(42.38)} & \textcolor{mytxtclr2}{(4.68)} & \textcolor{mytxtclr2}{(27.04)} & \textcolor{mytxtclr2}{(3.01)} & \textcolor{mytxtclr2}{(5.94)} & \textcolor{mytxtclr2}{(10.76)} & \textcolor{mytxtclr2}{(3.25)} & \textcolor{mytxtclr2}{(2.94)} & \textcolor{mytxtclr2}{(100.00)}\\
\hspace{1em}nwrk & 0.41 & 0.05 & 0.29 & 0.04 & 0.02 & 0.01 & 0.11 & 0.03 & 0.96\\
\textcolor{mytxtclr2}{\hspace{1em}} & \textcolor{mytxtclr2}{(42.18)} & \textcolor{mytxtclr2}{(5.01)} & \textcolor{mytxtclr2}{(30.61)} & \textcolor{mytxtclr2}{(4.40)} & \textcolor{mytxtclr2}{(2.47)} & \textcolor{mytxtclr2}{(0.59)} & \textcolor{mytxtclr2}{(11.16)} & \textcolor{mytxtclr2}{(3.59)} & \textcolor{mytxtclr2}{(100.00)}\\
\hspace{1em}ALL & 26.23 & 9.31 & 32.14 & 6.38 & 22.04 & 0.15 & 0.80 & 2.94 & 100.00\\
\textcolor{mytxtclr2}{\hspace{1em}} & \textcolor{mytxtclr2}{(26.23)} & \textcolor{mytxtclr2}{(9.31)} & \textcolor{mytxtclr2}{(32.14)} & \textcolor{mytxtclr2}{(6.38)} & \textcolor{mytxtclr2}{(22.04)} & \textcolor{mytxtclr2}{(0.15)} & \textcolor{mytxtclr2}{(0.80)} & \textcolor{mytxtclr2}{(2.94)} & \textcolor{mytxtclr2}{(100.00)}\\
\bottomrule
\end{tabular}}
\end{table}

\end{knitrout}

Both the row totals and column totals in the table are a measure of the percentage of population in each labour market state, the row totals being the measure of individuals in the state in quarter $t$ and the column totals being a measure of individuals in quarter $t+1$. The two totals do not exactly match for two reasons. First, if we number our quarters from $1$ to $N$ then the row totals are an average over quarters $1$ to $N-1$ while the column totals are an average over quarters $2$ to $N$. Second, the set of individuals included in the calculation between quarter $t$ to $t+1$ is not the same as the set of individuals included in the calculation between quarter $t+1$ and $t+2$ because of households entering and leaving the survey and because of attrition.

\subsection{Stylised facts}
Table~\ref{tab:transmat} presents us with a number of stylised facts. 

\begin{enumerate}

\item \textit{Changes of state are infrequent}. In the two tables most of the flows are concentrated along the diagonal. On an average, in each quarter, only 6.32\% of women and 10.68\% of men change their labour market state. This has a very important implication for measurement. Sources of measurement error or attrition bias which may be small with respect to the overall sample size may be large enough to swamp the small number of observations of state transitions. Therefore a much higher level of accuracy of observation is required for surveys meant to study transitions compared to surveys meant to study cross-sectional features. It is therefore worrying that the PLFS does not include mechanisms like reinterviews of a subset of respondents a short while after the original interview to quantify measurement errors.

\item \textit{Gross flows are much larger than net flows}. Even though gross flows are small relative to the size of the population, they are much larger than net flows. This indeed is the basic fact motivating the study of gross labour market flows, reconfirmed in our case. For example, on an average in every quarter 0.10\% of women moved from salaried employment to unemployment while at the same time 0.10\% of women moved back from unemployment to salaried employment. Thus on the net there is no flow from salaried employment to unemployment, though every quarter there is a significant number of women moving between these two employment states. These changes at the individual level which have obvious welfare implications would be completely missed by a study of net flows.

\item \textit{Labour market conditions are very different for men and women.} The low occupancy in employment states for women reconfirms the low labour force participation rate for women in India on which there is already an extensive literature. But the tables bring out the fact that this substantial difference in stocks is accompanied also by a substantial difference in flows in the sense that women experience much large inflows and outflows into employment. For example, the transition probability for women in salaried employment to move into unemployment or nonparticipation is $1.02+5.72=6.74$\% whereas for men the corresponding transition probability is $1.52+1.04=2.56$\%. These high proportionate rates of outflows for women are matched by equally high rates of inflows. Similar comparisons hold for casual work and self-employment. Thus, while women participate less in employment at any given moment of time, there is a much greater flux of women moving into and out of employment. This matches the results of~\citet{sarkar2019employment} and \citet{deshpande21} who also find that Indian women move into and out of employment at a significantly higher rate than Indian men.

\item \textit{Unemployment and non-participation overlap.} For women, flows into all three categories of employment is higher from non-participation than from unemployment. This could be explained by the very high percentage of women in the non-participation state. However, even for men the flow from non-participation to employment is substantial. For example, 0.30\% of the male population moved from non-participation to salaried employment which is not negligible compare to the 0.49\% of the male population that moved from unemployment to salaried employment. 

These direct movements from non-participation to employment could have two sources. One, it would include workers who complete their job searches between two observations and hence are never observed in the searching state. But, equally important is the possibility of response or measurement errors. Because of the stigma associated with unemployment some of the unemployed may respond that they are not looking for work. Or the responses of some of the non-employed may be incorrectly coded. In any case, our results show that it would be inappropriate to focus excessively on the unemployment rate as a measure of the functioning of the labour market. Among those recorded as non-participating there are many who do wish to participate.

\item \textit{Job-to-job flows are significant.} There are significant gross flows between the three employment states of salaried employment, casual employment and self-employment. So, for example, compared to 0.49\% of the male population flowing from unemployment to salaried work each quarter, 0.35\% flow from casual work and 0.58\% flow from self-employment each quarter. Thus these categories of work do not exist in ironclad compartments and employment-to-employment flows are as important as non-employment-to-employment flows. In fact we know from the studies in countries where matched employer-employee records are available that there exists substantial job-to-job flows. We cannot capture these flows in our data since we have no employer identifiers, but movement between the different kinds of employment gives us a glimpse of these flows. Also to the extent that salaried work is associated more with the formal sector and self-employment and casual work are associated more with the informal sector, these flows show that there exists significant worker porosity between these sectors of the urban Indian economy.

\item \textit{`nwrk' requires further disaggregation} The PLFS uses the employment status codes 62 and 72 for individuals who had work in household enterprises or salaried employment respectively but did not work due to reasons other than sickness. We aggregate these two codes into our `nwrk` state. There are significant flows to this state from employment states, specially from salaried employment. For example, for women the flow from salaried employment to `nwrk' is 0.17\% of women while the corresponding flow from salaried employment to unemployment is lower at 0.1\%.

While part of these flows would be persons taking a break or vacation, it is quite possible that they also include persons temporarily laid off or not being able to work due to disruptions in production.  

\begin{figure}[t]
\caption{\label{fig:earn-break}Earnings before and during `nwrk'}

\begin{knitrout}
\definecolor{shadecolor}{rgb}{0.969, 0.969, 0.969}\color{fgcolor}
\includegraphics[width=\textwidth]{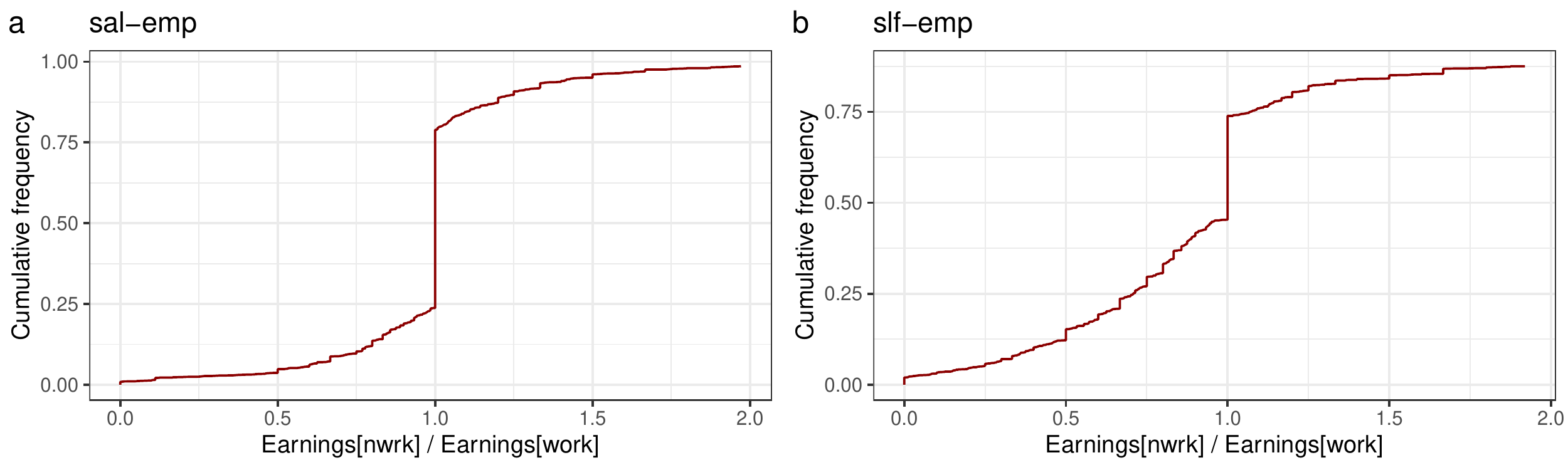} 
\end{knitrout}
\end{figure}

To unpack this further we look at individuals who transition from either salaried employment or self-employment to the state `nwrk' and compare their earnings before and after the transition. Figure~\ref{fig:earn-break} presents the empirical cumulative distribution function of the ratio of earnings after to earnings before entering the `nwrk' state. There is a large jump in both cases at 1, showing that for a large fraction---around 50\% for salaried employment and 25\% for self-employed---there is no change in earnings on moving to this state, so the transition is likely to be a voluntary break from work. For around 25\% in both cases there is actually a reported increase in income.\footnote{The self-employment category includes unpaid family workers whose incomes are recorded as zero and whose income can only increase when they move to another state.} And for around 25\% in the case of salaried employees and around 50\% in the case of the self-employed there is a decrease in income which might indicate some sort of involuntary inactivity. 

A look at the PLFS data for the post-COVID period shows a substantial increase in flows into `nwrk' and in the proportion of individuals earning nothing after entering this state. Given these indications of heterogeneity it would be useful if future versions of PLFS were to unpack this category further.

\end{enumerate}

\section{Diversity in flow rates by industry and region}\label{sec:ind-region}

To compare the labour market transition rates by industry and regions, we use the following definitions, following~\citet{davis1992gross},

\begin{align*}
  E_{i,t} &= \text {set of workers in labour market state $i$ at time $t$}\\
  \#S &= \text{number of elements in the set $S$ }\\
  \text{Entry rate}_{i,t} &=
    \frac{\#(E_{i,t+1} \cap E_{i,t}^c)}
    {(\#E_{i,t}+\#E_{i,t+1})/2}\\
  \text{Exit rate}_{i,t} &=
    \frac{\#(E_{i,t} \cap E_{i,t+1}^c)}
    {(\#E_{i,t}+\#E_{i,t+1})/2}\\
  \text{Gross flow}_{i,t} &= \text{Entry rate}_{i,t}+\text{Exit rate}_{i,t}
\end{align*}

The definitions of entry rate and exit rates differ from the usual definition of growth rates only in using the average of the occupancy in time $t$ and $t+1$ in the denominator, instead of the occupancy only at time $t$. The advantage of these alternative definition is that they produce numbers bounded between $0$ and $2$, making comparisons easier.

\begin{figure}
\caption{\label{fig:by-ind}Industry-level diversity (Entry and exit from industry)}
\begin{knitrout}
\definecolor{shadecolor}{rgb}{0.969, 0.969, 0.969}\color{fgcolor}

{\centering \includegraphics[width=\textwidth]{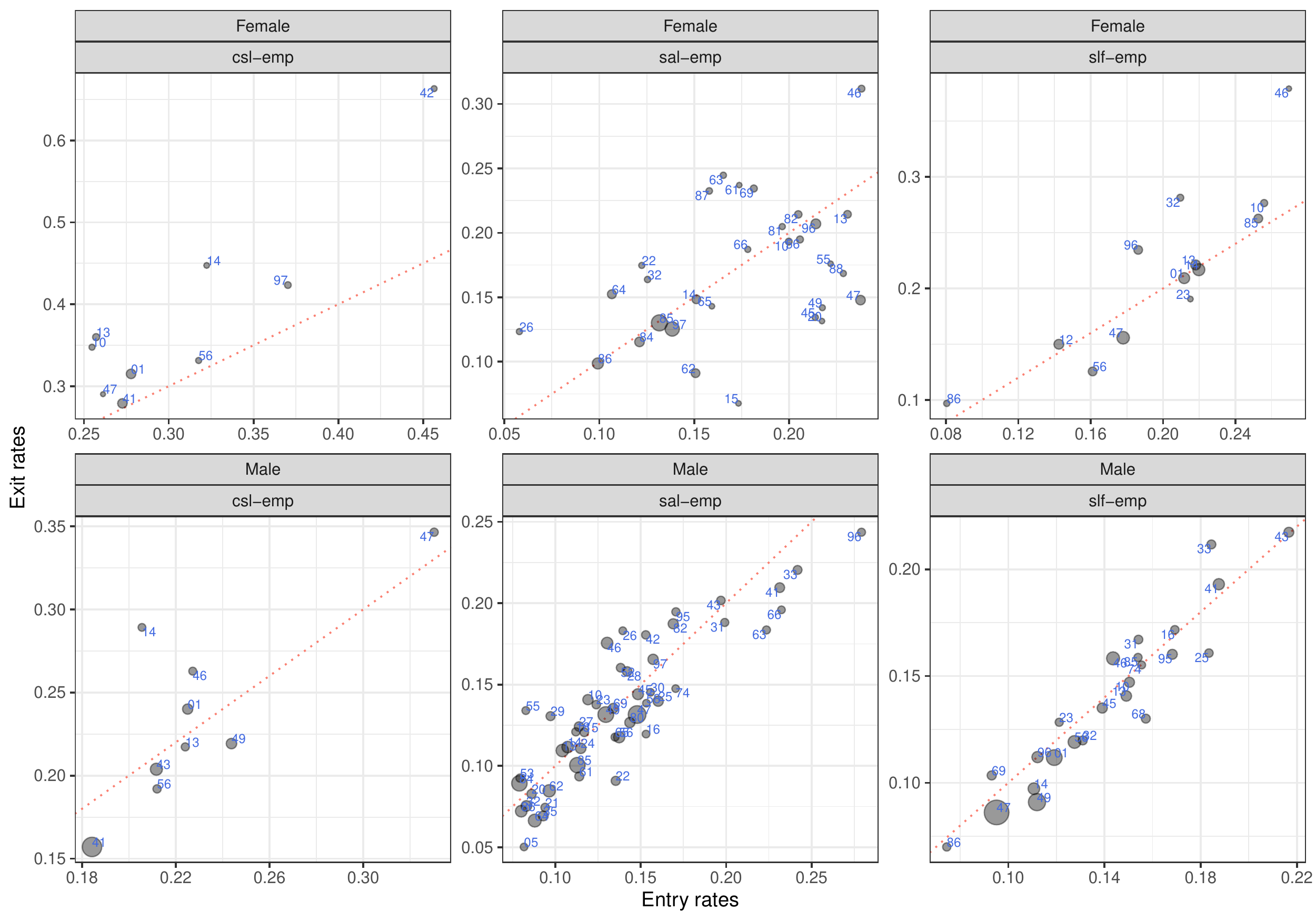} 

}

\end{knitrout}
\end{figure}

\begin{figure}
\caption{\label{fig:by-state}Regional diversity (Entry and exit from work)}
\begin{knitrout}
\definecolor{shadecolor}{rgb}{0.969, 0.969, 0.969}\color{fgcolor}

{\centering \includegraphics[width=\textwidth]{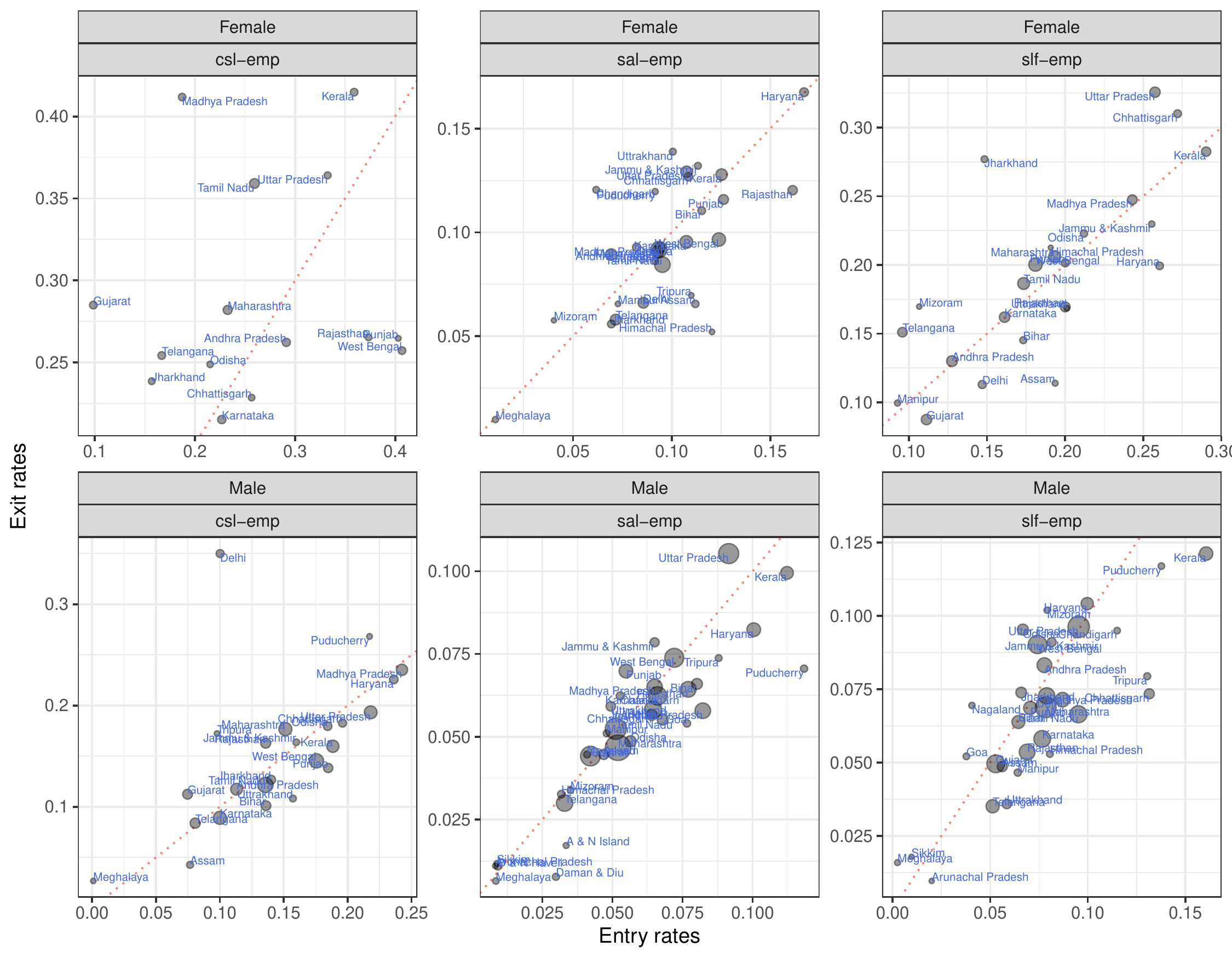} 

}

\end{knitrout}
\end{figure}

Figure~\ref{fig:by-ind} shows the average all-India entry and exit rates for two-digit NIC 2008 industry codes \citep{nic08}. Figure~\ref{fig:by-state} shows the average entry and exit rates from employment for the different states (provinces) of India. In both cases data for men and women and for casual-, salaried- and self-employment are plotted separately.

The first observation from these figures is the high degree of clustering in most cases around the dashed 45-degree line. Industries and states with high entry rates also tend to have high exit rates. Thus, the variation of entry and exit rates across states and industries is not due to the net expansion or shrinkage of industries. Rather it is due to the intrinsic labour market conditions in each industry.

The second observation is that the variation between industries and states is quite large. Even if one looks at a narrowly defined group such as salaried male workers, entry and exit rates range from 0.1 to 0.25 across 2-digit industries and 0.01 to 0.1 for entry and exit from work in different states.

One difficulty in interpreting the two figures separately is that the industry-wise composition of employment differs from state to state. To the extent that labour market transition rates are influenced by both state-specific and industry-specific factors, looking at each separately is subject to confounding. An industry might show high turnover because it is primarily situated in a state with weak labour market protections. On the other hand a state might show a high turnover because most of its workers are employed in industries where hiring and firing is cheap.

To try and decompose the industry and state effects, we estimate the following linear model which imposes an additive structure on these effects:

  \begin{align*}
    \text{[Gross flow rate]}
    &=\sum_{i,j,k} \beta_{ijk} D^I_i D^T_j D^G_k\\
    &+\sum_{l,k} \gamma_{lk} D^S_l D^G_k + \text{[Period dummies]}
    \end{align*}

Here we use the gross flow rate---the sum of entry and exit rates---as a dependent variable that captures the rate of labour turnover. On the right-hand side are a range of dummy variables
\begin{enumerate}
\item $D^I_i$: industry dummies, where $i$ ranges over 2-digit industry codes.
\item $D^T_j$: employment type dummies, where $j$ ranges over $\{\text{slf-emp},\text{csl-emp},\text{sal-emp}\}$.
\item $D^G_k$: gender dummies, where $k$ ranges over $\{Male,Female\}$.
\item $D^S_l$: state dummies, where $l$ ranges over the states of India
\end{enumerate}

The regression is fit to observations of gross flow rates in each quarter for each state,industry, employment type, and gender combination. To avoid noisy data from combinations with very few sample observations, we only keep those combinations whose total sampling multiplier is greater than $5\times 10^7$ for women and $10^8$ for men.

The regression is estimated without an intercept and therefore coefficients for all possible industry- employment type-gender combinations can be estimated. For the state-gender combinations, Maharashtra-Male is taken as the baseline.

\begin{knitrout}
\definecolor{shadecolor}{rgb}{0.969, 0.969, 0.969}\color{fgcolor}\begin{table}

\caption{\label{tab:coeff-ind}Regression coefficients, industry and employment type, $\beta_{ijk}$ (only coefficients significant at 5\% level of significance, baseline 0, equation estimated without intercept)}
\centering
\resizebox{\linewidth}{!}{
\fontsize{9}{11}\selectfont
\begin{tabular}[t]{>{\raggedright\arraybackslash}p{20em}llrr}
\toprule
\multicolumn{1}{c}{} & \multicolumn{1}{c}{} & \multicolumn{1}{c}{} & \multicolumn{2}{c}{Coefficient} \\
\cmidrule(l{3pt}r{3pt}){4-5}
Industry & NIC Code & Emp.\ Status & Female & Male\\
\midrule
Manufacture of chemical and chemical products & 20 & sal-emp &  & 0.26\\
Food and beverage service activities & 56 & sal-emp &  & 0.27\\
Manufacture of food products & 10 & slf-emp &  & 0.27\\
Wholesale trade, except of motor vehicles & 46 & sal-emp &  & 0.30\\
Land transport and transport via pipelines & 49 & csl-emp &  & 0.33\\
\addlinespace
Other manufacturing & 32 & sal-emp &  & 0.34\\
Land transport and transport via pipelines & 49 & slf-emp &  & 0.34\\
Wholesale trade, except of motor vehicles & 46 & slf-emp &  & 0.39\\
Computer programming etc & 62 & sal-emp & 0.47 & 0.32\\
Security and investigation agencies & 80 & sal-emp &  & 0.40\\
\addlinespace
Public administration and defence & 84 & sal-emp & 0.50 & 0.30\\
Education & 85 & sal-emp & 0.47 & 0.34\\
Land transport and transport via pipelines & 49 & sal-emp &  & 0.41\\
Manufacture of wearing apparel & 14 & sal-emp & 0.55 & 0.29\\
Manufacture of textiles & 13 & sal-emp & 0.49 & 0.34\\
\addlinespace
Food and beverage service activities & 56 & slf-emp & 0.49 & 0.36\\
Human health activities & 86 & sal-emp & 0.43 & \\
Activities of households as employers of domestic personnel & 97 & sal-emp & 0.46 & 0.41\\
Other personal service activities & 96 & sal-emp & 0.44 & \\
Retail trade, except of motor vehicles & 47 & slf-emp & 0.54 & 0.33\\
\addlinespace
Crop and animal production & 01 & slf-emp & 0.52 & 0.35\\
Manufacture of wearing apparel & 14 & slf-emp & 0.65 & 0.27\\
Financial services, except insurance and pension & 64 & sal-emp & 0.63 & 0.30\\
Manufacture of tobacco products & 12 & slf-emp & 0.46 & \\
Retail trade, except of motor vehicles & 47 & sal-emp & 0.55 & 0.43\\
\addlinespace
Specialized construction activities & 43 & csl-emp &  & 0.50\\
Construction of buildings & 41 & slf-emp &  & 0.51\\
Manufacture of textiles & 13 & slf-emp & 0.57 & 0.44\\
Crop and animal production & 01 & csl-emp & 0.72 & 0.30\\
Construction of buildings & 41 & csl-emp & 0.60 & 0.49\\
\addlinespace
Other personal service activities & 96 & slf-emp & 0.86 & 0.33\\
Education & 85 & slf-emp & 0.64 & \\
Manufacture of textiles & 13 & csl-emp & 1.00 & \\
Manufacture of rubber and plastic products & 22 & sal-emp &  & 1.06\\
\bottomrule
\end{tabular}}
\end{table}

\end{knitrout}

\begin{knitrout}
\definecolor{shadecolor}{rgb}{0.969, 0.969, 0.969}\color{fgcolor}\begin{table}

\caption{\label{tab:coeff-state}Regression coefficients, state $\gamma_{lk}$ (only coefficients significant at 5\% level of significance, baseline Maharashtra-Male)}
\centering
\fontsize{10}{12}\selectfont
\begin{tabular}[t]{lrr}
\toprule
State & Female & Male\\
\midrule
Telangana & -0.16 & -0.08\\
Karnataka & -0.11 & \\
Gujarat & -0.09 & -0.05\\
Maharashtra & -0.08 & \\
Andhra Pradesh & -0.08 & -0.06\\
Tamil Nadu & -0.07 & \\
Madhya Pradesh &  & 0.04\\
Uttar Pradesh &  & 0.06\\
Haryana &  & 0.09\\
Kerala &  & 0.09\\
\bottomrule
\end{tabular}
\end{table}

\end{knitrout}

Tables~\ref{tab:coeff-ind} and~\ref{tab:coeff-state} give the statistically significant coefficients for industry, employment type, gender combinations and state, gender combinations respectively.

Even after controlling for state effects, the results of Table~\ref{tab:coeff-ind} shows a wide variation in gross flow rates among industries and forms of employment, with coefficients ranging from 0.26 for male salaried workers in chemical manufacturing to 1.0 for female casual workers in the manufacture of textiles. Even within the same industry, workers with different forms of employment face different conditions. For example in education, salaried women  have a gross flow estimate of 0.47 while self-employed women have a gross flow estimate of 0.64.

When it come to states, the most significant fact about Table~\ref{tab:coeff-state} is the small number of entries. For most state-gender combinations the estimated coefficient is not statistically significant, confirming that most of the variation among states is driven by differences in industrial composition among states and not state-specific factors.

Unfortunately, there is no discernable pattern among the states whose coefficients are in fact statistically significant. We must conclude that these differences must be explained by idiosyncratic factors.

\section{Determinants of job loss and gain}\label{sec:gain-loss}

The analysis in the previous sections looked at gross flows between consecutive pairs of quarters. However, our data actually tracks workers for up to four quarters. In this section we use this additional information to look at how labour market states of workers are correlated over longer time spans. At the same time we extend our study of gender differences and see how other worker characteristics influence labour market transitions for men and women.

To keep the analysis tractable we collapse the labour market states into just two: employment and non-employment. We choose not to use the traditional three-way classification of employment, unemployment and non-participation in the light of our observation earlier on the possible conceptual and measurement overlap between unemployment and non-participation. 

To track the transitions between these two states we define two dummy variables \texttt{Lost} and \texttt{Gained}. \texttt{Lost} is $1$ for observations in which the individual is employed in that quarter but is not employed in the next quarter, and $0$ otherwise. \texttt{Gained} is $1$ for observations in which the individual is not employed in that quarter but is employed in the next quarter. Thus \texttt{Gained} and \texttt{Lost} measure movements into and out of employment. For the interpretation of the results to follow it must be remembered that both these variables have the value $0$ both for those who are out of employment and remain out of employment and for those who are in employment and remain in employment.

As explanatory variables we use the following worker characteristics:
\begin{enumerate}
\item \texttt{Very.Young}: Has the value $1$ if the individual's age is less than 21 and $0$ otherwise.
\item \texttt{Young}: Has the value $1$ if the individual's age is between 21 and 30 and $0$ otherwise.
\item \texttt{Graduate}: Has the value $1$ if the individual's general education level is graduation or higher and $0$ otherwise.
\item \texttt{Has.Child}: Has the value $1$ if the individual belongs to a family which has a child less than $5$ years of age and $0$ otherwise. We have to use this proxy for child-bearing since the data does not provide any information to link parents to children.
\item \texttt{Married}: Has the value $1$ if the individual is currently married and $0$ otherwise.
\end{enumerate}

We also include as explanatory variables the following two summaries of the individual's employment history:
\begin{enumerate}
\item \texttt{E.Ratio}: It is the net fraction of periods in which the individual has been observed to be employed. For worker~$i$ in quarter~$t$, let $E_{it}$ be the total number of quarters (not necessarily consecutive) up to and including $t$ in which we have observed them to be employed and $N_{it}$ be the total number of quarters (not necessarily consecutive) up to and including $t$ in which we have observed them not to be employed. Let $V_{it}$ be the visit number (between 1 and 4) for that individual in quarter $t$. Then we define:
\[\text{E.ratio}_{it} = \frac{E_{it}-N_{it}}{V_{it}}\]

\item \texttt{EN.Streak} For individual~$i$ in quarter~$t$, this is the number of consecutive periods up to and including $t$ for which they have been in the same state that they are in quarter~$t$. So for an employed individual this is the number of consecutive periods we have observed them to be employed while for non-employed individuals this is the number of consecutive periods we have observed them not in employment. This is set up as a categorical variable.
\end{enumerate}

Using these variables we estimate the following equations using logit.
  
  \begin{align*}
  \text{Lost/Gained}=1&+\text{Very.Young}+\text{Young}+\text{Graduate}\\
  &+\text{Has.Child}+\text{Married}+\text{E.Ratio}+\text{EN.Streak}\\
  &+\text{[Period dummies]}+\text{[Visit no. dummies]}
  \end{align*}

The equation for \texttt{Lost} is run on the set of individuals in employment while the equation on \texttt{Gained} is run on the set of individuals not in of employment. For each individual, data from visits upto one less than the last recorded visit is used since the variables \texttt{Lost} and \texttt{Gained} are defined by looking one quarter ahead. Observations for all quarters and all workers are pooled together. Each model is estimated separately for men and women.

\begin{table}
\caption{Determinants of job gain and loss. Average marginal effects from logit model.}
\begin{center}
\begin{small}
\begin{tabular}{l D{.}{.}{2.6} D{.}{.}{2.6} D{.}{.}{2.6} D{.}{.}{2.6}}
\toprule
 & \multicolumn{1}{c}{Gained (Female)} & \multicolumn{1}{c}{Gained (Male)} & \multicolumn{1}{c}{Lost (Female)} & \multicolumn{1}{c}{Lost (Male)} \\
\midrule
Very.Young  & -0.015^{***} & -0.054^{***} & 0.077^{***}  & 0.027^{***}  \\
            & (0.001)      & (0.004)      & (0.013)      & (0.003)      \\
Young       & 0.001        & 0.019^{***}  & 0.022^{***}  & 0.007^{***}  \\
            & (0.001)      & (0.004)      & (0.005)      & (0.002)      \\
Graduate    & -0.003^{**}  & -0.028^{***} & -0.043^{***} & -0.012^{***} \\
            & (0.001)      & (0.002)      & (0.004)      & (0.001)      \\
Has.Child   & -0.005^{***} & 0.014^{***}  & 0.009        & 0.002        \\
            & (0.001)      & (0.003)      & (0.005)      & (0.001)      \\
Married     & -0.007^{***} & 0.037^{***}  & 0.036^{***}  & -0.026^{***} \\
            & (0.001)      & (0.005)      & (0.004)      & (0.002)      \\
E.Ratio     & 0.052^{***}  & 0.120^{***}  & -0.089^{***} & -0.039^{***} \\
            & (0.003)      & (0.008)      & (0.012)      & (0.004)      \\
EN.Streak=2 & -0.014^{***} & -0.030^{***} & -0.065^{***} & -0.021^{***} \\
            & (0.003)      & (0.007)      & (0.012)      & (0.005)      \\
EN.Streak=3 & -0.020^{***} & -0.046^{***} & -0.099^{***} & -0.034^{***} \\
            & (0.004)      & (0.011)      & (0.015)      & (0.006)      \\
\bottomrule
\multicolumn{5}{l}{\tiny{$^{***}p<0.001$; $^{**}p<0.01$; $^{*}p<0.05$}}
\end{tabular}
\end{small}
\label{tab:logit-gainloss}
\end{center}
\end{table}

Table~\ref{tab:logit-gainloss} gives the average marginal effects from the logit models. The numbers in parentheses are standard errors.

Being `very young', i.e.\ being in the age group 15--20, increases the probability of job losses and decreases the probability of job gains. There are two potential explanations for this. One would be the precarity of the job market for young, inexperienced and less-educated workers. The other would be the fact that this is the age range for school and undergraduate education, and individuals in educational institutions in these age groups are more likely to continue education than join employment and those in employment are likely to leave employment to continue their education. Indeed, in our sample, 81.1\% of the `very young' who are not in employment report attending educational institutions. Of those leaving employment in this age group, 27.55\% report attending an educational institution in the next quarter. Thus education does have a role to play. However, even if the equations for \texttt{Lost} are restimated after dropping observations with employment to education transitions, being `very young' still has a positive, significant effect, showing that precarity plays a role too.

For the young, i.e.\ the age group 21--30, the probabilities for both job loss and job gain are higher than the baseline. In this age group the outcomes for men and women diverge significantly. The additional loss probability for women is now more than three times that for men. The additional gain probability is positive and significant for men and very small and statistically insignificant for women. In this age group while men are more likely to enter employment, presumably after completing their education, women are losing employment. 

Being a graduate reduces the probability of both job loss and job gain for both men and women. This is presumably because those with graduate and higher degrees have a more secure work life. Those in employment are less likely to lose jobs, and those out of employment are there by choice and not because of unwanted job loss, and hence are less likely to move into employment.

Having a child in the family does not have a statistically significant effect on the probability of job loss. The effect on job gain is siginificant and in opposite directions for men and women: negative (though small in magnitude) for women and positive for men.

The effect of marriage too is quite different for men and women. Being married reduces the probability of job loss and increases the probability of job gain for men. For women it is the opposite: the probability of job loss goes up and that of job gain goes down.

Overall, the picture that emerges is that of a gender- and age-differentiated labour market with women and the young facing greater employment instability.

When it comes to work histories the \texttt{EN.Streak} variables are negative and significant in all regressions. To recall, these measure that number of consecutive prior quarter in which the individual has been in the same state as in the quarter of observation. Thus this variable has a different meaning in the \texttt{Gained} and \texttt{Lost} regressions. For the \texttt{Gained} equation, estimated on the sample of individuals not in employment, it is the number of consecutive prior quarters they have not been in employment. For the \texttt{Lost} equation, estimated for the sample of individuals in employment, it is the number of consecutive prior quarters they have been in employment. Thus the results show the existence of inertia in job market states. Prolonged periods of non-employment perpetuate non-employment and prolonged periods of employment perpetuate employment. The magnitude of the estimated effects is larger for streaks of length three compared with streaks of length two, showing that this inertia becomes stronger the longer the worker is in employment or non-employment.

Even after controlling for the streak variables, the variable \texttt{E.Ratio} remains statistically and economically significant: negative in job loss regressions and positive in job gain regressions. Thus it is not only consecutive spans of employment and non-employment which affect the probabilities of job gains and losses. Past occurrences of employment make getting back into employment more likely and getting out of employment less likely, even after controlling for the immediately prior history.

\section{Conclusion}
This study reveals a rich world of micro-level transitions in the urban Indian labour market with a great degree of diversity across genders, industries and regions. It is hoped that as more data becomes available we will be able to better tease out the institutional factors underlying this diversity. Essential to that endeavour would be panel data on jobs from the firm side and the ability to match workers and employers. The introduction of PLFS gives us hope that Indian official data would expand to cover these aspects as well.

\clearpage
\bibliography{plfs.bib}
\end{document}